\documentclass[aps,prl,onecolumn,showpacs]{revtex4-1}
\usepackage{amsmath, amssymb}
\usepackage{amsfonts}
\usepackage{graphicx}
\usepackage{bm}
\usepackage{color}
\begin{document}

\title{Crack front dynamics: the interplay of singular geometry and crack instabilities }
\author{Itamar Kolvin, Gil Cohen and Jay Fineberg}
%\thanks{}
\affiliation{The Racah Institute of Physics, The Hebrew University of Jerusalem, Jerusalem, Israel 91000}

\date{\today}

\begin{abstract}
 When fast cracks become unstable to microscopic branching (micro-branching), fracture no longer occurs in an effective 2D medium. We follow in-plane crack front dynamics via real-time measurements in brittle gels as micro-branching unfolds and progresses. We first show that {\em spatially local} energy balance quantitatively describes crack dynamics, even when translational invariance is badly broken. Furthermore, our results explain micro-branch dynamics; why micro-branches form along spatially localized chains and how finite-time formation of cusps along the crack front leads to their death.
\end{abstract}

\maketitle
Although fracture mechanics is a mature field, many of its simplest questions remain unanswered. Cracks govern material stability.  The classic 2D descriptions of fracture \cite{Freund.90} shows that the existence of a crack focuses elastic energy stored in a strained material into a single point - the crack tip. There, stresses diverge as $\sigma\sim K/\sqrt{r}$ where $r$ is the distance from the tip and $K$ is a coefficient called the stress intensity factor. The motion of the crack is then governed by the physical demand of energy balance; that $G$, the elastic energy flowing to the crack tip, equals the fracture energy $\Gamma$ (the energy dissipated per unit area of crack extension). For example, a sudden increase in $\Gamma$ will cause a corresponding decrease in crack velocity, since part of the elastic energy that is used to separate the crack faces  is now used to compensate for the increase in fracture energy.  Energy balance is the basis for the continuum description of the dynamics of simple, straight cracks in brittle materials.

Cracks, however, fracture 3D materials.  Whereas a crack's tip in 2D materials is a singular {\em point}, in 3D the leading edge of a crack forms a singular {\em line}, the {\em crack front}.
 In ``simple" cracks, the crack front is a straight line with no overt dynamics. For this reason, 2D descriptions of fracture dynamics for simple cracks are sufficient and the classical 2D theory delineated above \cite{Freund.90} describes crack dynamics perfectly \cite{ BFM10,Sharon.99,Goldman.2010}, as long as they propagate along a straight-line trajectory.

Simple cracks are generically unstable entities \cite{Fineberg99instabilityin} that leave complex 3D surfaces in their wake \cite{hull1999fractography, baumberger.2013,Scheibert.2010,Guerra.2012}. Their intrinsic 3D character calls for a study of the dynamics of crack {\em fronts}  \cite{Ramanathan.97, Morrissey.00,Leblond.11,Adda-Bedia.13, Pons.10}.  Even when a crack is constrained to a plane, the geometry of the crack front is important. For example, the advance of quasi-static fronts in patterned materials is influenced by the long-range elastic forces that act along crack fronts  \cite{ DalmasVandembroucq2009, Chopin2011,Patinet.2013}. A recent study \cite{Ponson.2012} of quasi-static peeling of adhesive tape demonstrates the striking consequences of manipulating front geometry to  enhance material toughness.  The elasticity of crack fronts is also important to explain the fluctuating dynamics of quasi-static planar cracks propagating through random heterogeneities \cite{ santucci.2010,bonamy2008crackling, Roux.2013}.

\begin{figure}[ht!]
\includegraphics[scale=0.34]{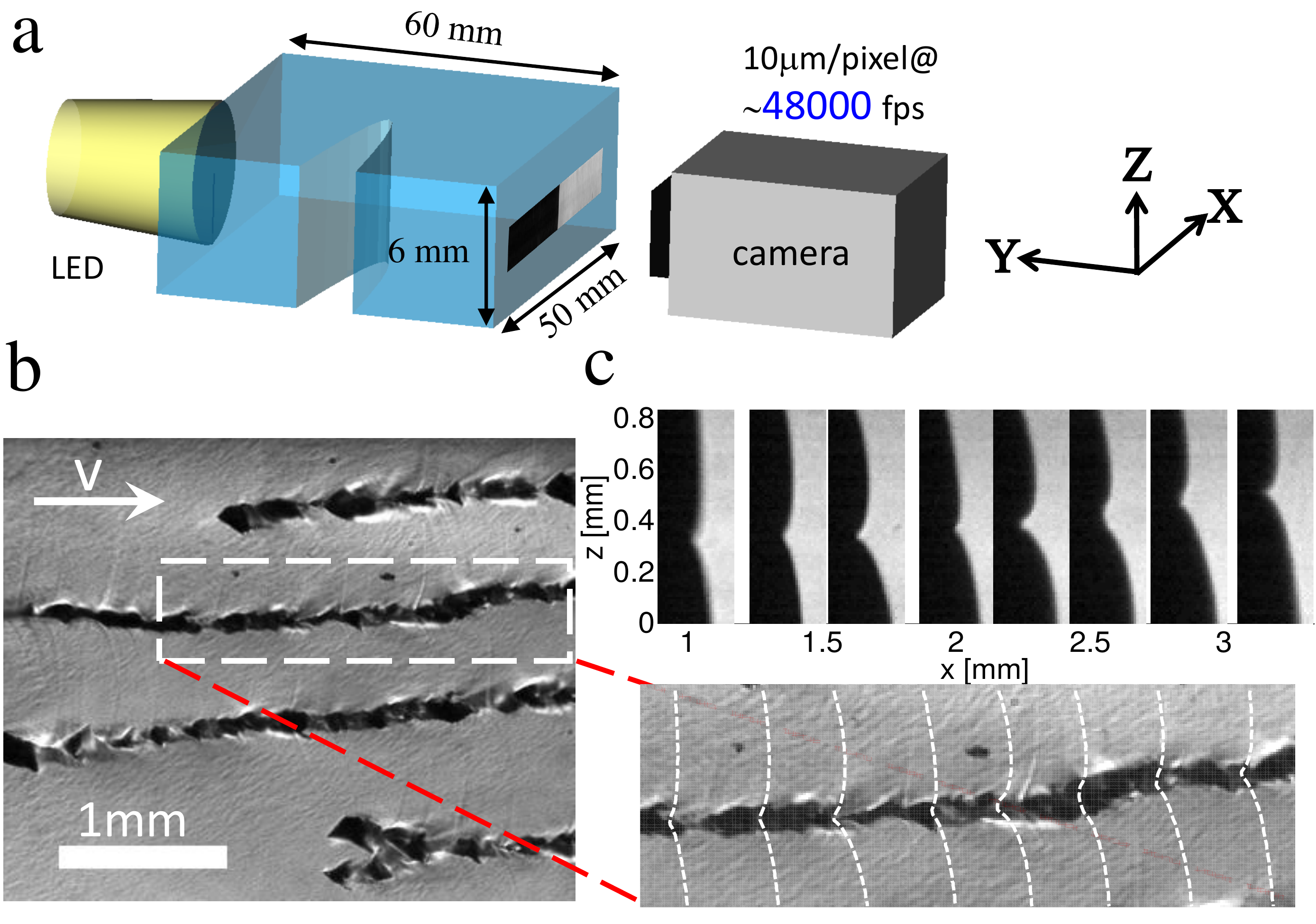}
\caption{(a) The experimental setup. A rectangular block of polyacrylamide gel is strained by displacement of its $Y$ boundaries.
Crack fronts, propagating within the sample's mid-plane are visualized via shadowgraph with a high-speed camera; illumination is via collimated LED light normal to the fracture ($XZ$) plane.
(b) A \textit{post-mortem} fracture surface section containing four nearly parallel directed chains of micro-branches (``branch-lines").  (c)  {\em Top.} The imaged fronts that formed the branch-line denoted by the box in (b). {\em Bottom.} A close up photograph of the corresponding fracture surface with overlaid edge-detected fronts.}  \label{Fig1}
\end{figure}

When cracks are rapid, experiments suggest \cite{sharon2002crack} that distortions of a crack front produce inertia in the ``massless" cracks described by 2D theory; an intrinsic 3D effect. The micro-branching instability of rapid simple cracks is an example where a crack front loses its symmetry {\em dynamically}.  Simple cracks in brittle materials experience this instability above a critical crack velocity $v_c\sim 0.3c_R$, where $c_R$ is the material's Rayleigh wave speed \cite{Fineberg99instabilityin,sharon2002crack,Livne.05}. For velocities $v>v_c$, directed chains (``{\em branch-lines}") of micron-scale branched cracks are spontaneously generated (see Fig. \ref{Fig1}b). Each ``micro-branch" branches away from the main crack to propagate beneath its fracture surface.  Every micro-branch creates additional fracture surface and hence, increases the value of $\Gamma$ that is felt by the main crack \cite{Sharon.96}. As, for short times, the total energy flux to the crack front is constant, micro-branching also reduces the energy flux to the rest of the crack. A single micro-branch may effectively increase $\Gamma$ by up to 100\%  (if it creates an additional  crack propagating parallel to the main crack).  One may therefore consider micro-branches as energy sinks that are dynamically ``toggled" on and off.   As micro-branches are localized along the crack front, these perturbations are spatially localized in both parallel and transverse to the propagation direction. Therefore, when excited, micro-branches locally perturb initially straight crack fronts to produce large fluctuations in $v$.

In this Letter, we study the propagation of fast crack fronts during the micro-branching instability.
We will provide a detailed account of in-plane front dynamics as micro-branches nucleate, grow and eventually die.
This will provide us with new insights as to how front geometry and motion continuously feed each other; eventually conspiring to cause micro-branch ``death".

We study crack front dynamics by performing real-time visualization of the fronts as they traverse the fracture surface.
We do this by using brittle polyacrylamide gels composed of a 13.8\% (w/v) of acrylamide and a 1:37.5 (w/w) bisacrylamide-to-acrylamide ratio where $c_R=5.2$ m/s.
When scaled by $c_R$, these gels exhibit both the same single crack dynamics  \cite{Goldman.2010} and micro-branching phenomenology \cite{Livne.05} as more conventional brittle materials such as glass and PMMA.
The advantage of using gels is in reducing the wave speeds, hence crack velocities, by three orders of magnitude.

Our experimental system is schematically described in Fig. \ref{Fig1}(a). Our gel samples are cast to be rectangular blocks of dimensions $50\times 60\times 6 mm^3$ $(X\times Y\times Z)$, where $X$, $Y$, and $Z$ are, respectively, the propagation, loading and thickness directions.
We image the crack front by shining collimated LED light through the samples in a direction normal to the XZ (fracture) plane.
 The strong curvature at the crack tip deflects the light at the crack front, creating a sharp shadow boundary in the image plane and producing a 2D projection of the crack front.
We capture front dynamics by imaging the front using high-speed (IDT-Y4) camera at $\sim\! 48000$ frames/sec in a ``$X\!\times \!Z$" window of $10\!\times\! 1\,mm^2$ located at the center section of the gel sample.
Our spatial resolution was $\sim\! 10$ micron per pixel.

We generated crack front dynamics in the micro-branching regime by displacing the system's boundaries by a 10-12\% strain, prior to initiating fracture. Fracture was initiated by inserting a small ``seed" crack at the sample's edge, midway between its vertical boundaries.
 This procedure produced crack velocities of 0.1-0.5$c_R$ along the mid-plane of the sample.
All velocities, $v(z)$, referred to in this paper are the {\em normal} velocities to the front at each point $z$.

In Fig. 1(b) we present a typical picture of the fracture surface created. Crack propagation is complex, forming micro-branches at several locations along the front. The four branch-lines appearing in the figure are oriented parallel to the mean crack velocity. To characterize the corresponding front dynamics, we focus on the formation of a single branch-line, denoted by the dashed box in Fig. \ref{Fig1}(b). In Fig. \ref{Fig1}(c) we present a series of snapshots of successive crack fronts within this box, together with a close up of the resulting \textit{post-mortem} branch-line formed by these fronts. A direct comparison between the fronts and resulting fracture surface (Fig. \ref{Fig1}(c), bottom) shows that a one-to-one correspondence exists between micro-branches on the fracture surface and loci of high curvature along the front.
\begin{figure}[h]
\includegraphics[scale=0.5]{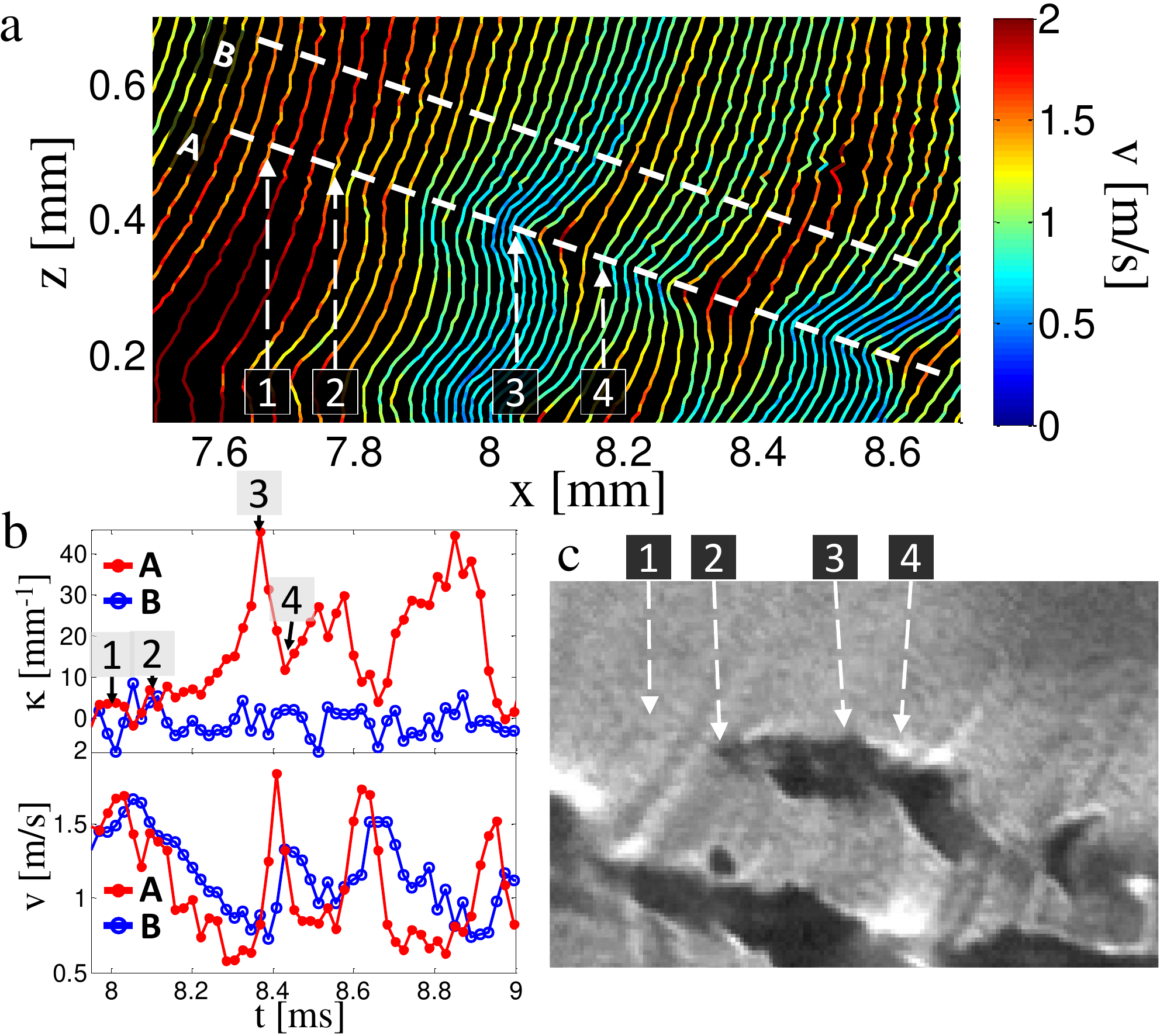}
\caption{Three consecutive branching events along a branch-line. (a)
The series of fronts captured during branch-line formation. Colors indicate local velocity levels. Marked points show (\textbf{1}) an unperturbed front (\textbf{2}) local velocity decrease associated with micro-branch nucleation (\textbf{3}) maximum local curvature (\textbf{4}) formation of a new micro-branch.  (b) Plots of local curvature ({\em Top}) and local velocity ({\em Bottom}) along the cuts \textbf{A} (dots) and \textbf{B} (open symbols) in (a).
(c) The corresponding post-mortem showing the branch-line associated with the fronts in (a). }\label{Fig2}
\end{figure}

Fig. \ref{Fig2} presents a detailed picture of the motion within a typical section of a front that includes three well-defined micro-branching events.
Along the crest of a single branch-line marked \textbf{A} we can see that, prior to the nucleation of the first micro-branch, (\textbf{1}) the front has nearly zero curvature.
Then, with the nucleation of the first micro-branch (\textbf{2}) the local front velocity along \textbf{A} gradually decreases.
The apparent cause for the slowing down is the effective increase in local fracture energy due to the incipient micro-branching.
The resulting gradient in velocity along the front produces an increase of front curvature.
The front acquires a locally concave shape which becomes more and more pronounced until (\textbf{3}) the curvature suddenly drops as the local velocity peaks to $\sim 100\%$ above the mean front velocity.
Examination of the fracture surface suggests that the release of curvature and the velocity jump are coincident with the ``death" of the micro-branch.
Immediately afterwards (\textbf{4}) the velocity drops again as a new micro-branch nucleates.
In contrast to the strong fluctuations along \textbf{A}, the ``quiet" adjacent mirror-like region \cite{mirrorfootnote} marked \textbf{B} shows practically no change in curvature.
We do, however, see significant variations in velocity which, after a delay, follow the changes in velocity along the branch-line \textbf{A} .

Let us consider these results.
First, the association of micro-branch nucleation and death with local velocity decrease
and increase along the branch-line is qualitatively consistent with {\em local} energy balance; $G=\Gamma$.
Our second observation is that after the front acquires curvature and the micro-branch dies,
the local velocity does not only return to the mean velocity level, but doubles it.
These observations suggest that front curvature must be increasing $G$ locally to produce these high velocity peaks. This idea is supported by a well-known result derived by Rice \cite{rice1985first} for {\em static} fronts.
Rice computed the \textit{static} stress contribution made by a small in-plane perturbation to a straight front.  When a front $x = x_0+\delta x(z)$ is perturbed in the sense that $|dx/dz|\ll 1$, the stress maintains the inverse-square-root singularity at each point along the front,
but $K = K_0+\delta K(z)$ varies along the front as
\begin{equation}
\label{Rice}
\frac{\delta K(z)}{K_0} = \frac{1}{2\pi}PV\int\! \frac{\delta x(z')-\delta x(z) }{(z'-z)^2}\, dz'\,,
\end{equation}
where $xz$ is the fracture plane, $x$ is the crack propagation direction and the unperturbed front is $x(z)=x_0$ with a stress intensity factor $K_0$.
A salient feature of this integral is that it is solely determined by the front geometry.
It generates a stabilizing force (``line tension") that tends to restore the front to a flat configuration, since  $G\propto K^2$ or $\delta G/G_0 = 2\delta K/K_0$. This force is countered by the local increase of the fracture energy, $\delta \Gamma(z)$,  generated by the micro-branch. In general, we cannot measure $\delta \Gamma$ directly. When, however, a micro-branch ``dies", $\delta \Gamma\sim 0$ and the elastic energy stored in the front curvature locally accelerates the crack. Evidence for this is seen Fig. \ref{Fig2}b along the \textbf{A} line, where we see that each peak in curvature is followed by a peak in local front velocity as the line tension is released.

%Since $G\propto K^2$ or $\delta G/G_0 = 2\delta K/K_0$,
%front curvature stores elastic energy which, we believe, is responsible for the high velocity values %observed when a micro-branch dies and the local fracture energy drops.
%To summarize, front dynamics appear to be governed by three inter-dependent players: the local %velocity, the local line tension and the local (effective) fracture energy. At least two of those are %needed to predict further front propagation.

\begin{figure}[h]
\includegraphics[scale=0.7]{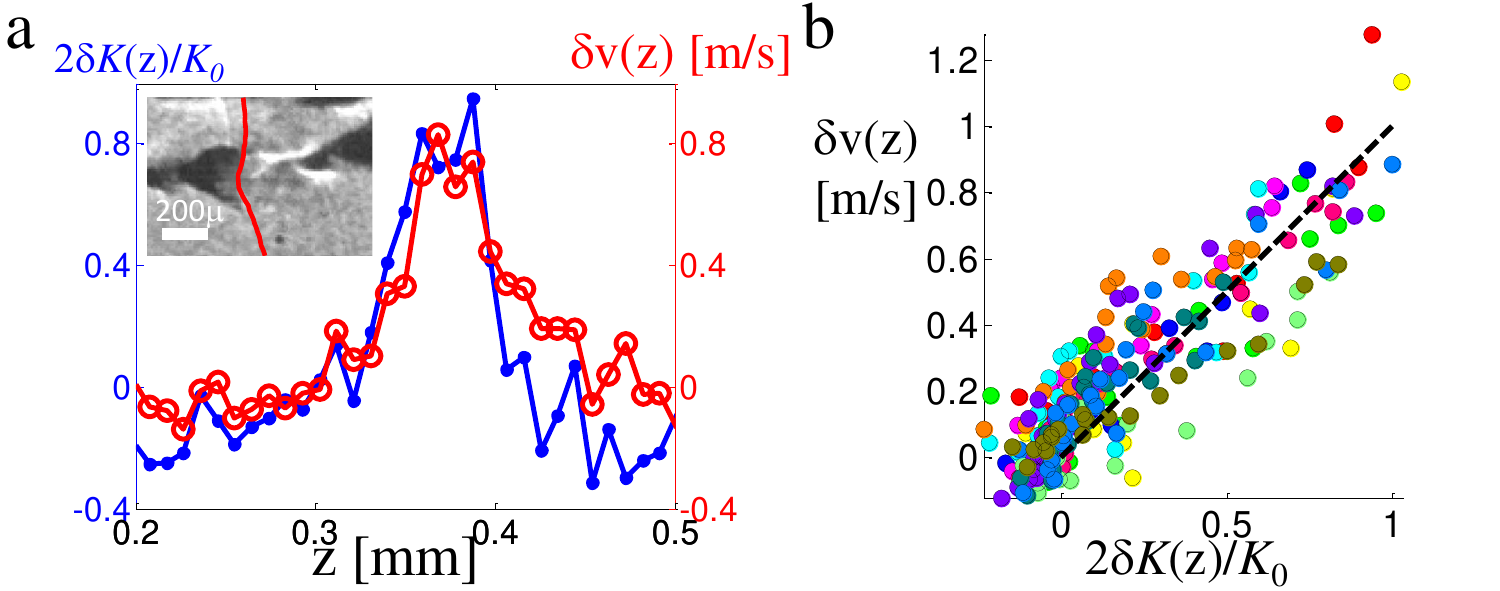}
\caption{Comparison of the instantaneous velocity and local stress intensity factor, computed by Eq. (\ref{Rice}), along a front at the moment of micro-branch ``death". (a) Spatial stress (dots) and velocity variations (open circles) for the front in the inset are well-correlated. ({\em inset}) Overlay of the front at the moment of maximal velocity on the resulting \textit{post-mortem} fracture surface. Front positioning is approximate.  (b) $\delta v(z)$ vs $\delta G/G_0 = 2\delta K/K_0$ for both the data in (a) and twelve such releases from other events. Different colors indicate different events. The dashed ``y=x" line is a guide to the eye.  }\label{Fig3}
\end{figure}

While these statements appear qualitatively accurate, are they quantitatively correct?
Let us now focus on the moment of micro-branch death; the moments corresponding to the peak in local front velocity at the instant where $\delta \Gamma \sim 0$. We consider thirteen cases, with mean crack velocities ranging between $0.1-0.3 c_R$,
where we identify a clear transition from a branched to smooth surface in the {\em post-mortem} image.

At these instances we can directly compare the relative velocity deviation from the mean $\delta v(z) = v(z) - \langle v(z) \rangle$  and the built-up energy release rate $\delta G/G_0 = 2\delta K(z)/K_0$ computed by Eq. (1).
% we should try to make the read area a bit simpler
As Eq. (1) was derived for fronts parallel to the $z$ axis, we computed $\delta K/K_0$ over symmetric regions of the front around the point of maximum curvature by first rotating the front to make edges of this region sit on an $x=const$ line. The average $\langle v(z) \rangle$ was computed for velocity values in the same region.

The result of the comparison is presented in Fig. \ref{Fig3}.  We see that the correction for the stress intensity factor given by Eq. (\ref{Rice}) is clearly correlated with the instantaneous velocity profile at the moment of micro-branch ``death". Despite being only a first-order correction, Eq. (1) is shown here to work quite well. This is demonstrated locally along a typical front in Fig. \ref{Fig3}a and in another twelve similar instances in Fig. \ref{Fig3}b. It is a manifest demonstration of how front curvature is translated into motion.
We note that we see no observable effect of the proximity of neighboring branch lines in this collapse.  This is consistent with the observations of \cite{ChopinThese} in quasi-static propagation.
Since we compare here a dimensionless quantity with velocity $\delta v = 2 v_0 \delta K/K_0$, the coefficient of proportionality $v_0$ must also have the dimensions of velocity. Its values $v_0 = 1\pm 0.3 m/s$ are far from the characteristic wave velocity ($c_R\sim 5 m/s$), but close to the measured mean crack velocities $\langle v\rangle = 0.7-1.5 m/s$. This might be related to the strong velocity dependence of the fracture energy in polyacrylamide gels \cite{Goldman.2010}. We give a possible origin of this coefficient in the supplementary material \cite{supplementary,GoldmanUnpublished}.

\begin{figure}
\includegraphics[scale=0.6]{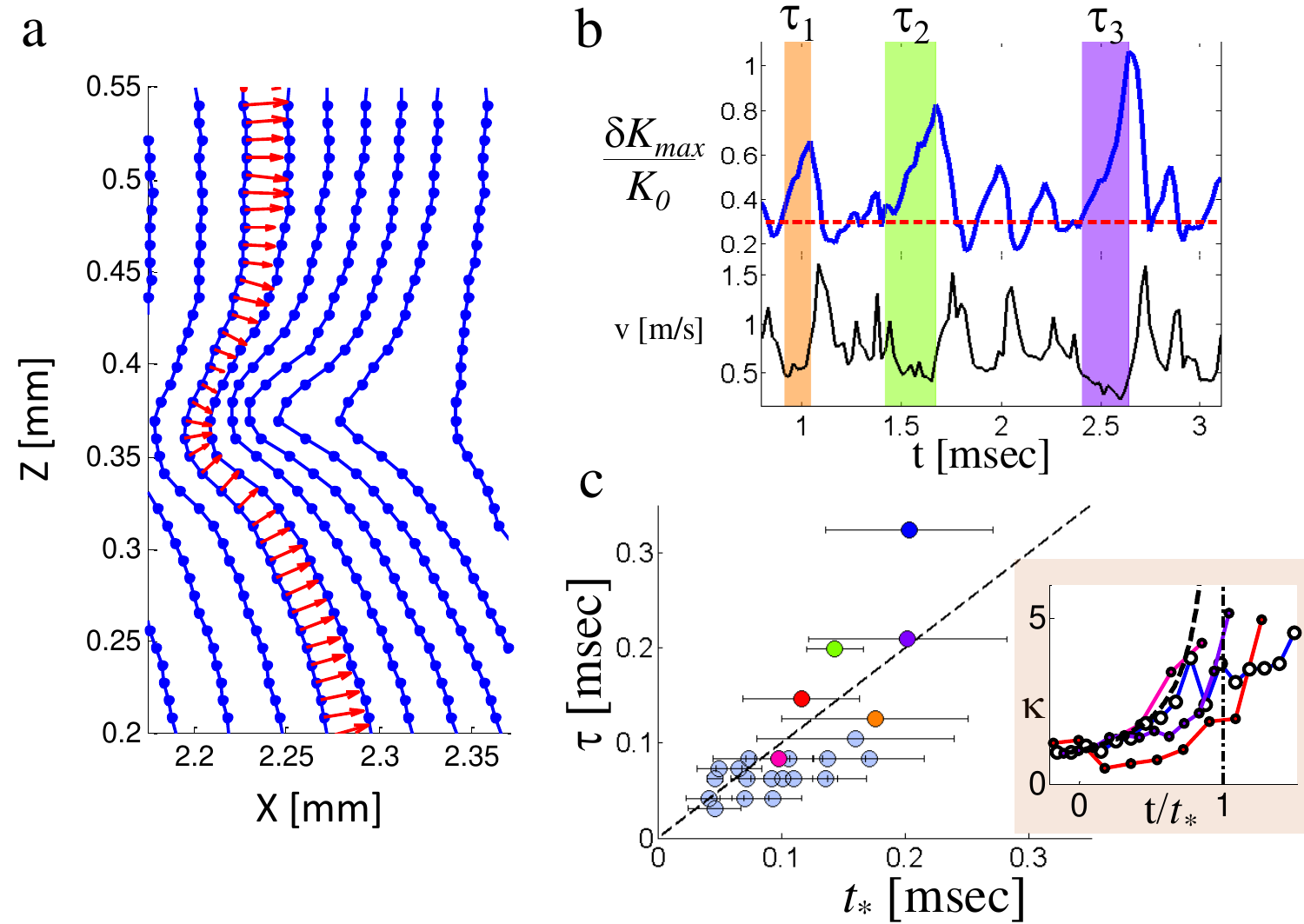}
\caption{(a) Cusp formation during a typical micro-branching event. We present every second front. Red arrows denote normal displacements. (b) Time-series of the peak value $\delta K/K_0$ (top) and front velocity (bottom) along the crest of a branch-line show a few cycles of build-up and release. We define the micro-branch lifetime, $\tau$, by the interval over which $\delta K/K_0$ first surpasses (micro-branch birth) and then drops below (micro-branch death) a threshold value $ 0.3$ (red line). Note that during stress build-up, the local velocity stays approximately constant. Shaded areas: 3 examples of $\tau$ for different branching events. (c) $\tau$ compared to predicted cusp formation times $t_* = 1/(\kappa_0 v_n)$.  $\kappa_0$ is the initial maximal curvature and $v_n$ is the initial normal velocity at the location of maximum curvature.  Colored points denote events in (b) and those shown in the inset. (\textit{inset}) Curvature blow-up dynamics. Time is normalized by $t_*$. $t=0$ corresponds to the initial front. $\kappa$ is the maximal curvature of each successive front normalized by $\kappa_0$.  Black dashed line: predicted finite-time blow-up.}
 \label{Fig4} \end{figure}

Let us now consider the front evolution depicted in Fig. 4 where the process leading to the build-up of the ``line tension" prior to the death of a micro-branch is highlighted. During the buildup of the line tension, $v$ along the crest of the branch-line is approximately constant during the rapid increase in curvature that leads to cusp formation.
  In Fig. 4(a) we present a series of fronts leading to a release event.
The release is preceded by a rapid increase in curvature which culminates when the
front attains a cusp-like form. As in the events depicted in Fig. 2, local increases in $v$ immediately follow cusp formation.

It is well-known that propagating fronts may develop shocks in curvature \cite{KPZ}, or cusps. If a cusp does form, we would expect the stress intensity factor to locally diverge according to Eq. (\ref{Rice}). It is, therefore, likely that this high stress could promote the ``death" of the micro-branch and hence trigger the local release of a front.

How can this picture be tested? A nearly constant normal velocity, $v(z,t)=v_n$, where $v_n$ is taken to be the initial velocity at the beginning of the event, coupled with an initially curved section of the front will spontaneously produce a cusp with no further assumptions.
To see this, we rewrite $v=v_n$ in terms of front coordinates; $\partial_t x = v_n \sqrt{1+(\partial_z x)^2}$.
Differentiating by $z$, we obtain an equation of motion for the local slope $u = -\partial_z x$:
\begin{equation}
\frac{\partial u}{\partial t}+v_n \frac{u}{\sqrt{1+u^2}}\frac{\partial u}{\partial z} = 0\,.
\end{equation}
\noindent It is known that this equation develops shocks (i.e. cusps in $u$) in finite-time
where $\partial_z^2 x = - \partial_z u  \sim (t_*-t)^{-1}$ and $t_* = 1/(\kappa_0 v_n)$ when $\kappa_0$ is the maximal front curvature at $t = 0$.
%This equation reduces to the inviscid Burgers equation in the
%limit of small slopes $|u|\ll 1$.

The constant velocity model provides us with a prediction for the time required for
the front to form a cusp $t_*$. Is this timescale related to the lifetime, $\tau$, of a micro-branching event? To test this, we consider an ensemble of micro-branching events. As depicted in Fig. 4(b), we define $\tau$ as the time between the initial stress build-up and stress release points for each event, using a threshold of $\delta K/K_0 =0.3$ to define these points. We use the initial maximal curvature, $\kappa_0$, and the normal velocity, $v_n$, to evaluate the predicted cusp formation time, $t_*$.
(We measure $\kappa_0$ by rotating the front to eliminate any mean slope and fitting the region of highest curvature with a parabola. $v_n$ is the average normal velocity over the same range in $z$.)  We consider only events where the estimated error in $t_*$ was no larger than 50\%.

As shown in Fig. 4(c) this simplistic model works rather well. Within the limits of our experimental resolution $\tau \simeq t_*$; over the
stress build-up we observe rapid growth of curvature that culminates near the predicted $t_*$. Moreover, our predicted $t_*$ indeed correspond to the death of each micro-branch.

This being said, micro-branch dynamics do not always proceed directly to curvature blow-up. In some cases the front may ``slip"; undergoing partial release during an event before final cusp formation. An example of such slip is the curve denoted by open symbols in the inset of Fig. 4 (c).

In conclusion, we have presented the first experimental exploration of rapid crack front in-plane dynamics. Our observation that micro-branches act as sinks that effectively increase the fracture energy agrees qualitatively with a local energy balance. The induced front curvature may explain the micro-branch localization in $z$ as the velocity overshoots upon micro-branch release appear to generate new micro-branches, thereby forming branch-lines. Moreover, we have shown that the pronounced velocity overshoot at the moment of micro-branch ``death" is well correlated with the static contribution for the stress intensity factor resulting from front curvature. This is a clear demonstration of how stresses distributed along a crack front translate locally into dynamics.

Nevertheless, the demonstrated validity of the expression (Eq. (\ref{Rice})) is intriguing for several reasons.
First, Eq. (1) is a first-order correction in the local slope of the front, while the fronts we considered contained mild slopes.
Our current data do not have the precision to investigate the effects of higher order corrections \cite{Katzav.2006,Leblond.2012,Vasoya.2014} and it would be interesting to test their implications in the future. Secondly, Eq. (1) was derived for static fronts while we consider moving fronts.
 In the work of Ramanathan \& Fisher \cite{Ramanathan.97} the static theory was expanded to include dynamic effects.
 These include waves that propagate on the front at velocity $\sim\! \sqrt{c_R^2-v^2}$.
Since in our experiments $v^2/c_R^2 \sim 0.1 \ll 1$ and because of the small size of micro-branches ( $\sim\! 100\mu$) and relatively long life-time ($\sim\! 100\mu s$) we expect any signature of the waves to average during micro-branch growth.
We might expect the agreement with  Eq. (1) to break down at larger $v/c_R$, when such inertial effects become important.
Lastly, apart from the increase in fracture area caused by the out-of-plane motions of the front, micro-branches should also influence the local energy release rate by introducing shearing stresses.
The combined effects of these contributions on dynamics demand further study.

In addition, while recent experiments \cite{Goldman.2015} provide a mechanism for micro-branch birth, we have seen that the front dynamics induced by this ``birth" could lead to eventual micro-branch ``death"; front curvature caused by micro-branch initiation may lead to micro-branch death via the large line tension generated by dynamic cusp formation. This scenario could provide a key to understanding why micro-branches remain small and do not develop.
The above results are all purely 3D effects that underline the necessity of extending fracture mechanics to the third dimension.

\textit{Acknowledgements.} The authors acknowledge the support of the European Research Council (Grant No. 267256) and Israel Science Foundation (Grant 76/11). We thank Baruch Meerson for helpful discussions on cusp formation in fronts.
\bibliography{bibl}

\end{document}

\widetext
\begin{center}
\textbf{\large Supplemental Material for Crack Front Dynamics}
\end{center}

\section{A heuristic calculation of $v_0$}

From Fig. 3 of the main text we concluded that at the moment of micro-branch ``death", the local velocity is proportional to the line tension given by Eq. (1). 
The origin of the proportionality constant, $v_0$, needs still to be accounted for. Here we suggest a possible heuristic explanation. 

Our statring point is the energy balance between the energy release rate and the fracture energy
\begin{equation}
G=\Gamma\,.
\end{equation}
We proceed by locally perturbing both sides of the balance to take into account the variation of $\Gamma$ and $G$ with local front geometry and velocity.
First, we note that $G\propto g(v) K^2$ where $g(v)\simeq 1-v/c_R$ is the dynamical correction factor of $G$ \cite{Freund.90} and $K$ here is the non-dynamic part of the stress intensity factor. 
Then perturbing around a straight front configuration with a constant energy release rate $G=G_0$ we have 
\begin{equation}
\frac{\delta G}{G_0} = \frac{d\log g}{dv}\delta v(z)+2\frac{\delta K(z)}{K_0}\,,
\end{equation}
where $\delta K(z)/K_0$ is given by Eq. (1).
Next, let's discuss the fracture energy. In polyacrylamide gels the fracture energy for a simple crack is a strong function of velocity \cite{Goldman.2010}.
We will term it here $\Gamma_0(v)$. 
Since in the instances we consider in Fig.3 there is no contribution from the micro-branch to the fracture energy, the local perturbation is simply
\begin{equation}
\frac{\delta\Gamma}{\Gamma_0} = \frac{d\log\Gamma_0}{dv}\delta v\,.
\end{equation}
Equating $\delta G/G_0 = \delta\Gamma/\Gamma_0$ we find that $\delta v = 2 v_0\delta K/K_0$ 
with 
\begin{equation}
v_0 = \left[\frac{d\log\Gamma_0}{dv}-\frac{d\log g}{dv}\right]^{-1}\,.
\end{equation}
To evaluate the expression for $v_0$ we use the results of \cite{Goldman.2010,GoldmanUnpublished}.
We approximate the fracture energy by a linear function $\Gamma_0(v)\simeq \Gamma(v_m) +\Gamma_0'(v_m)(v-v_m)$ around $v_m = 0.3c_R$. 
Then $d\log\Gamma_0/dv = 1/(v+u)$ where $u = \Gamma_0(v_m)/\Gamma_0'(v_m)-v_m =(0.25\pm 0.1)c_R $. 
Noting that also $d\log g/dv \simeq -1/(c_R-v)$
we see that  $v_0 = (v+u)(1-v/c_R)/(1+u/c_R)$. Since the mean front velocities for the data presented in Fig. 3 vary in the range $ v=0.8-1.6m/s$ the calculated proportionality coefficient is then  $v_0 =1.4\pm 0.3 m/s$, which is within the margin of error of the value of $v_0$ obtained from Fig. 3.

\section{Validity of the constant velocity assumption during micro-branch growth}

In the main text we argued that during the stress build-up phase of a micro-branching event the velocity stays virtually constant in the region of the micro-branch. Here in Fig. S1 we present statistics of 37 events where we compute the normalized standard deviation of local velocities for each events. Events where velocity markedly deviates from the mean are rare. 
\begin{figure}[hb!]
\includegraphics[scale=1]{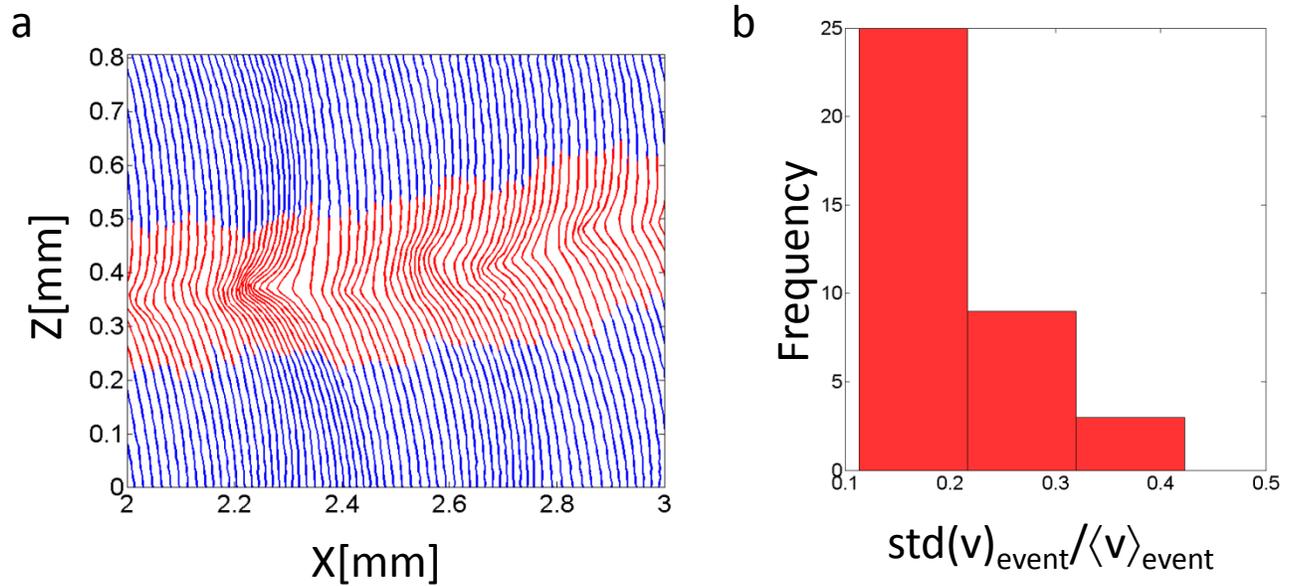}
\caption{The constant velocity assumption  for the cusp formation model. (a) An example of a series of consecutive crack fronts. In red are the "cusp" regions where the velocity statistics were taken from. (b) A  histogram of normalized standard deviation values for the 37 micro-branching events. For each event considered in Fig.4c of the main text, we gathered  velocity values from all the "cusp" regions of all the fronts, taking the fronts from the first passing the threshold to the one reaching maximum stress.
Then for each event we calculated the mean value of all velocity values $\langle v\rangle_{event}$ and used it to normalize the velocity values.}
\end{figure}

\bibliography{suppbibl}